\documentclass[proceedings]{JHEP3}

\PrHEP{PrHEP hep2001}			
\conference{International Europhysics Conference on HEP}		

\usepackage{epsf,epsfig}                   

\newcommand{\be}{\begin{equation}}
\newcommand{\ee}{\end{equation}}
\newcommand{\bea}{\begin{eqnarray}}
\newcommand{\eea}{\end{eqnarray}}

\def\bsigma{\mbox{\protect\boldmath $\sigma$}}
\newcommand{\reff}[1]{(\ref{#1})}

\def\gtapprox{\gtrsim}

\title{Discrete non-Abelian groups and asymptotically free models}

\author{Sergio Caracciolo\\
        Dipartimento di Fisica dell'Universit\`a di Milano, 
        Sezione INFN di Pisa and NEST-INFM, Italy\\
	E-mail: \email{Sergio.Caracciolo@sns.it}}

\author{Andrea Montanari\\
        Lab. de Physique Th\'eorique de l'Ecole Normale Sup\'erieure, 
        Paris, France\\
        E-mail: \email{Andrea.Montanari@lpt.ens.fr}}

\author{{Andrea Pelissetto}\\
       Dipartimento di Fisica dell'Universit\`a di Roma and 
       Sezione INFN di Roma, Italy \\
	E-mail: \email{Andrea.Pelissetto@roma1.infn.it}}

\abstract{We consider a two-dimensional $\sigma$-model with 
discrete icosahedral/dodecahedral symmetry. 
Using the perturbative renormalization group,
we argue that this model
has a different continuum limit with respect to the $O(3)$ $\sigma$ model.
Such an argument is confirmed by a high-precision numerical simulation.
}

\begin{document}


Recently, there has been interest in the critical behavior of 
two-dimensional $\sigma$-models in which the spins take values in some discrete 
subset of the sphere. In particular, two groups 
\cite{PS-98a,PS-98b,PS-01,HN-01a,HN-01b} studied 
the nearest-neighbor $\sigma$-model
\be
H = \beta \sum_{<ij>} \bsigma_i\cdot \bsigma_j,
\ee
in which the spins have unit length and belong to the vertices of a Platonic 
solid, i.e. of a tetrahedron, cube, octahedron, icosahedron, or dodecahedron. 
Several quantities 
have been computed: the renormalized two-point function, the 
current-current correlation function, 
the finite-size scaling (FSS) curve for the 
second-moment correlation length,  and the four-point renormalized 
coupling. Surprisingly enough, the results for the icosahedral and the 
dodecahedral model are very close to the $O(3)$ ones, suggesting that 
these three models might have the same continuum limit. 
Patrascioiu and Seiler \cite{PS-98a,PS-98b,PS-01} considered these 
results as evidence for the $O(3)$ $\sigma$-model not
being asymptotically free, since the discrete-symmetry models have 
a finite $\beta$ 
phase transition, which cannot be described in perturbation theory.
However, the overwhelming evidence we have collected in the years in favor 
of asymptotic freedom made Hasenfratz and Niedermayer \cite{HN-01a,HN-01b} 
suggest 
that, may be, the icosahedral and the dodecahedral models have 
an asymptotically-free continuum limit, 
in spite of the fact that the critical point is at a finite value of $\beta$. 

Here, we wish to show that, by using some standard assumptions, the 
perturbative renormalization-group (RG) approach predicts that the 
suggestion of  Hasenfratz and Niedermayer cannot be true. 
If the continuum limit of the $O(3)$ $\sigma$-model is correctly 
described by the perturbative RG, then any discrete-symmetry model cannot
belong to the same universality class of the $O(3)$ $\sigma$-model. 

The argument goes as follows \cite{CMP-01}. Consider the Hamiltonian 
\be
H = \beta \sum_{<i,j>}\bsigma_i\cdot\bsigma_j - 
     h \sum_i I_n(\bsigma_i), 
\label{Hmisto}
\ee
where $\bsigma_i$ is an $O(3)$ unit spin and $I_n(\bsigma_i)$ 
is a polynomial in $\bsigma_i$ with the following properties:
it has $O(3)$ spin $n$; 
the maxima (or minima) of $I_n(\bsigma_x)$ correspond to the set 
of vertices of a Platonic solid;
it is invariant under the discrete-symmetry group of the solid.
For all Platonic solids,
it can be shown explicitly that such a polynomial exists.
The model \reff{Hmisto} interpolates between the $O(3)$ $(h=0)$ and 
the discrete-symmetry  model $(|h|=+\infty)$. 
Now, with quite standard assumptions, 
one can show that $I_n(\bsigma)$ is a {\em relevant perturbation} of the 
$O(3)$ fixed point. In other words, any arbitrarily small 
perturbation with discrete symmetry 
of the $O(3)$ $\sigma$-model drives the system to a 
different fixed point. 

The argument is fairly standard. Consider a $p$-point connected correlation 
function $G^{(p)}(\beta,h)$ at zero external momenta in a finite box $L^2$. 
If $hL^2\ll 1$ and  $\xi\gg L$, we can compute the correlation function 
in perturbation theory, obtaining
\be
G^{(p)}(\beta,h) = \sum_{i,j=0}^\infty t^i h^j\ a_{ij}^{(p)}(L),
\label{PT-GP}
\ee
where $t\equiv 1/\beta$.
The coefficients of the expansion diverge as $L\to \infty$, since the 
infinite-volume correlation function cannot be computed directly 
in perturbation theory because of infrared divergences. However,
by using the perturbative expansion \reff{PT-GP}, one can show that 
in the continuum limit $G^{(p)}(\beta,h)$ satisfies the RG equation
\be
\left[-a{\partial\over \partial a} + 
      W(t) {\partial\over \partial t} + 
      \gamma^{(n)}(t) h {\partial\over \partial h} + 
      {p\over2} \gamma(t) \right] G^{(p)}(\beta,h) = 0,
\label{RG-eq}
\ee
where $W(t)$, $\gamma^{(n)}(t)$, and $\gamma(t)$ are $L$-independent 
RG functions. Then, we make the following assumption: 
\begin{itemize}
\item[] The RG equation \reff{RG-eq}---but {\em not}
 the expansion \reff{PT-GP} we 
        started from---is valid for all values of $L$, including $L=\infty$.
\end{itemize}
Such an assumption is routinely made in the perturbative analysis of the 
$\sigma$-model and is used, for instance, to obtain the small-$t$ 
behavior of long-distance quantities, such  as the susceptibility, 
correlation length,
and so on. Solving Eq.~\reff{RG-eq}, we obtain in the infinite-volume limit
\be
G^{(p)}(\beta,h) = G^{(p)}(\beta,0) \Phi^{(p)}(z),
\label{sol-RG}
\ee
where 
\be
z \equiv h t^\rho \exp (4\pi/t) \sim h \xi(t)^2 [\log \xi(t)]^\sigma,
\label{def-z}
\ee
$\rho$ and $\sigma$ are universal exponents that can be easily computed 
by using the perturbative results of Ref. \cite{CP-94},
$\Phi^{(p)}(z)$ is a nonperturbative crossover function, and $\xi(t)$
is the correlation length for $h=0$. Precisely, Eq. \reff{sol-RG} is valid 
in the crossover limit $t\to 0$, $h\to 0$, keeping $z$ fixed. 
Equations \reff{sol-RG} and \reff{def-z} show that 
$I_n(\bsigma)$ is a relevant perturbation with RG eigenvalue $2$, as 
expected on the basis of dimensional analysis. Different physical 
results (i.e. different results for universal quantities) are obtained 
by varying the variable $z$, as usual in the vicinity of a point perturbed 
by two relevant perturbations (more precisely the thermal direction 
is marginally relevant).  
Thus, within the standard perturbative approach, the discrete-symmetry
model and the 
$O(3)$ model are expected to have different continuum limits. 

The previous argument together with the numerical results of
Refs. \cite{PS-98a,PS-98b,PS-01,HN-01a,HN-01b} puts asymptotic freedom 
on a dangerous ground 
since it shows that the conventional scenario is wrong if the icosahedral
or the dodecahedral models have 
the some continuum limit of the $O(3)$ model. We have thus decided 
to extend the previous numerical work and indeed, we have found good 
evidence that the $O(3)$ model and the discrete-symmetry model belong to 
different universality classes: the conventional scenario is saved. 
However, the surprising fact is that these differences appear only 
very near to the critical point, i.e. for $\xi_\infty \gtrsim 10^5$!

In the numerical simulation we have considered the Hamiltonian \reff{Hmisto}
with 
\bea
I_6(\bsigma) &=& \sigma^6_z - 5 \sigma_z^4\left(\sigma_x^2 + \sigma_y^2\right)
   + 5 \sigma_z^2 \left(\sigma_x^2 + \sigma_y^2\right)^2 
\nonumber \\
   && + 2 \sigma_x \sigma_z
     \left(\sigma_x^4 - 10 \sigma_x^2 \sigma_y^2 + \sigma_y^4\right),
\eea
and $h = 0.1$. Such a polynomial is invariant under the rotation group 
of the icosahedron and of the dodecahedron.
We measured the second-moment correlation length 
as defined in Refs. \cite{CEPS-95,CEFPS-95}, and the spin-$n$ 
susceptibilities
\be
\chi_n = \sum_x \left\langle P_n\left(\sigma_0\cdot\sigma_x\right)\right\rangle,
\ee
where $P_n(x)$ is a Legendre polynomial, for $n=1,3,4$.
For each observable ${\cal O}(L,\beta)$, 
we considered the so-called step function,
i.e. the ratio ${\cal O}(2L,\beta)/{\cal O}(L,\beta)$, which,
in the continuum limit should become a universal function of 
$\xi(L,\beta)/L$, i.e.
\be
{{\cal O}(2L,\beta)\over {\cal O}(L,\beta)} = 
   F_{\cal O}\left({\xi(L,\beta)\over L}\right) + O(L^{-\omega},\xi^{-\omega}).
\ee
We measured the step function of the above-mentioned  observables in the 
discrete-symmetry theory (i.e. keeping $h=0.1$ fixed) and in the $O(3)$ model,
thereby extending the results of Refs. \cite{CEPS-95,CEFPS-95}.
If the two models have the same continuum limit, the function computed 
for $h=0.1$ and $h=0$ should coincide.

\DOUBLEFIGURE[tpb]{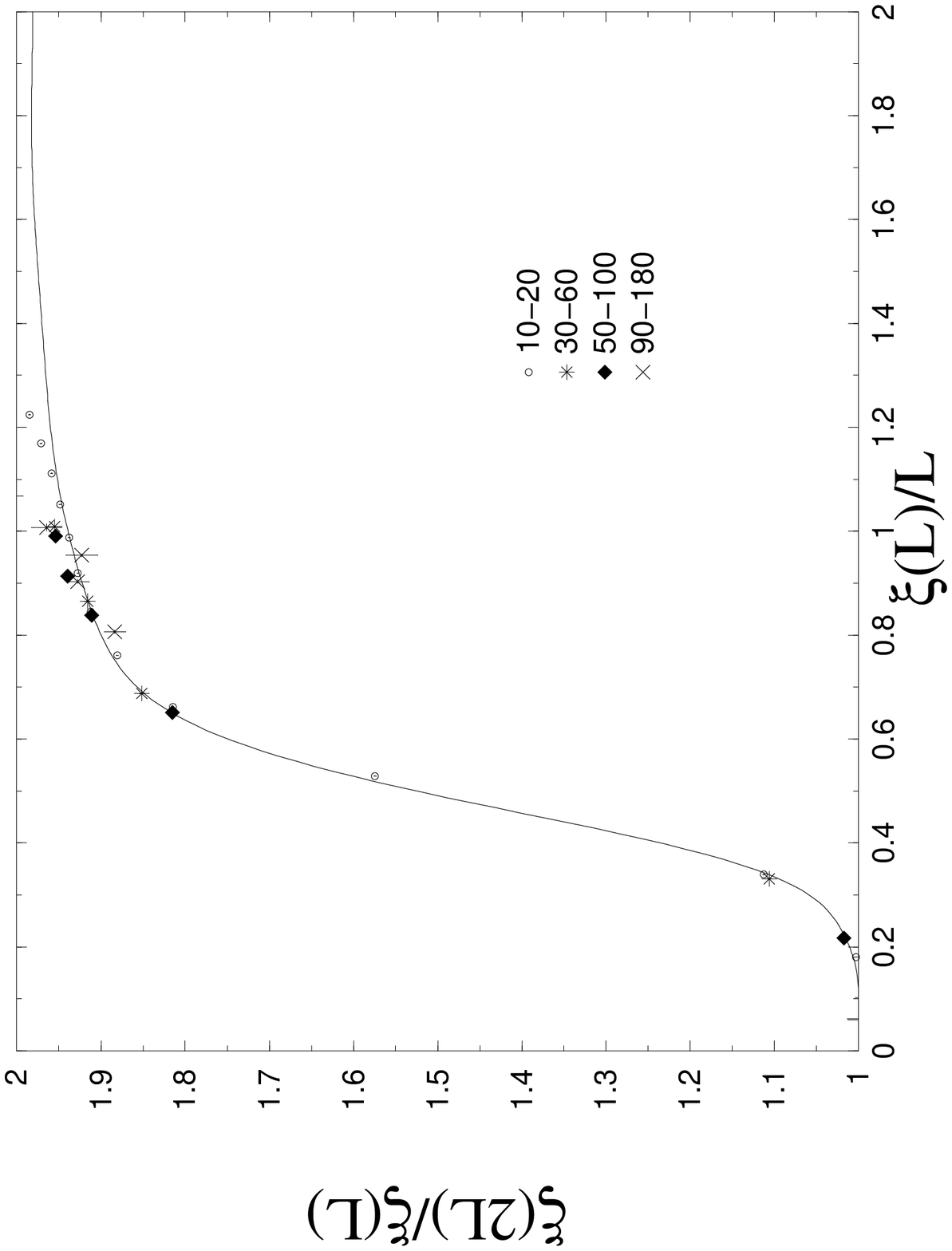,angle=-90,width=1.0\linewidth}%
{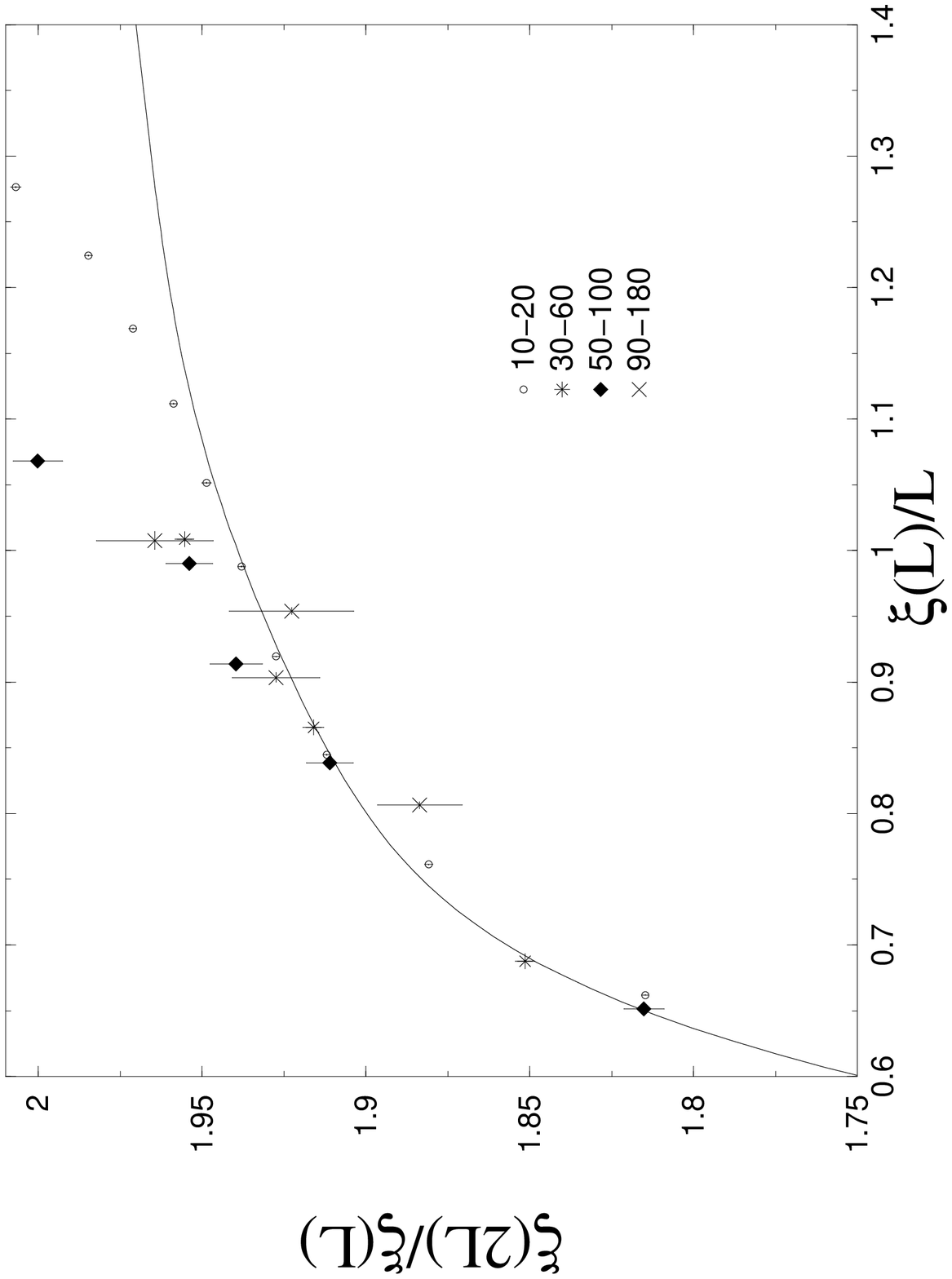,angle=-90,width=1.0\linewidth}%
{FSS function for the second-moment correlation length. \label{xi}}%
{FSS function for the second-moment correlation length. 
Here, we restrict the horizontal range to $0.6\le x \le 1.4$.\label{xibis}}

\DOUBLEFIGURE[tpb]{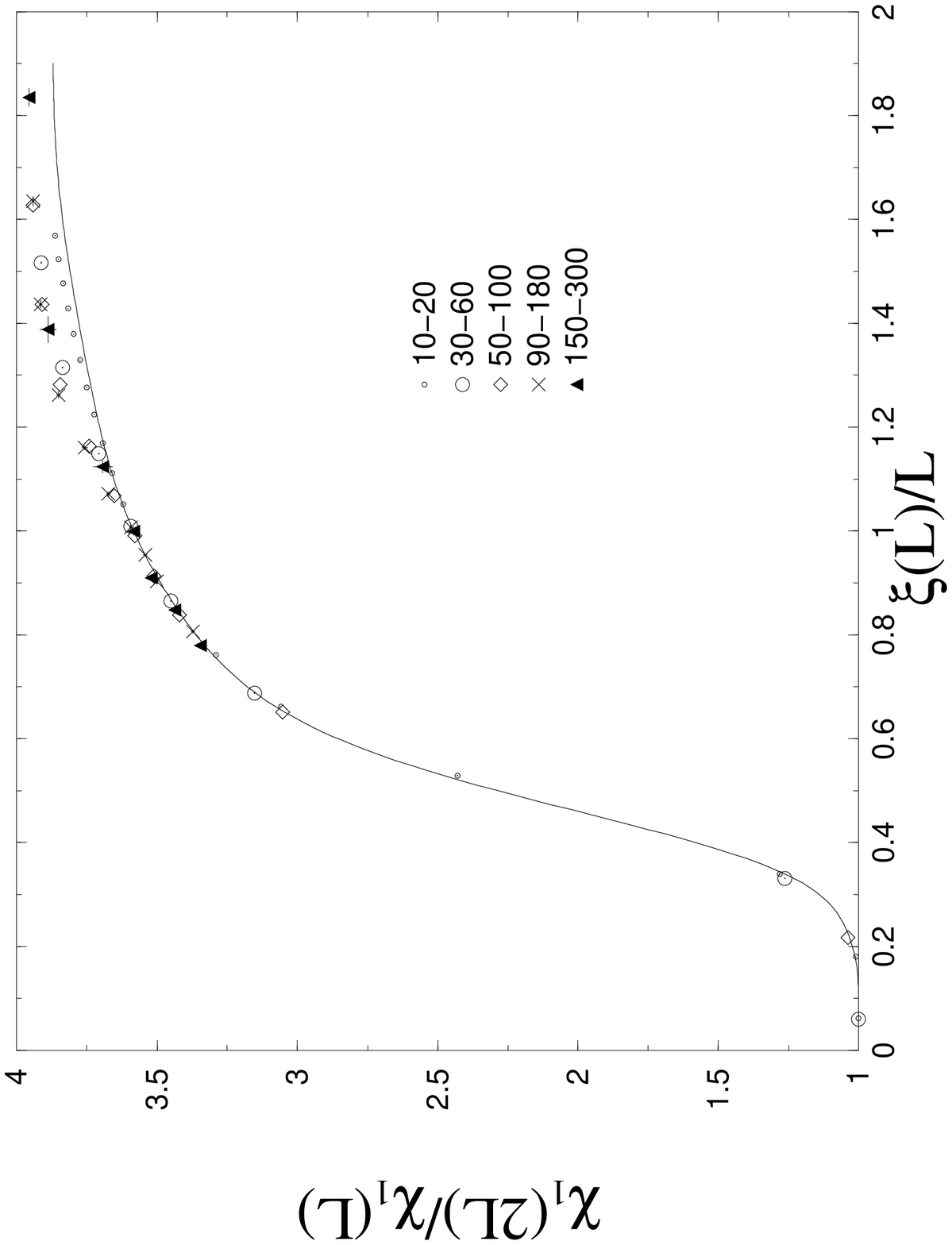,angle=-90,width=1.0\linewidth}%
{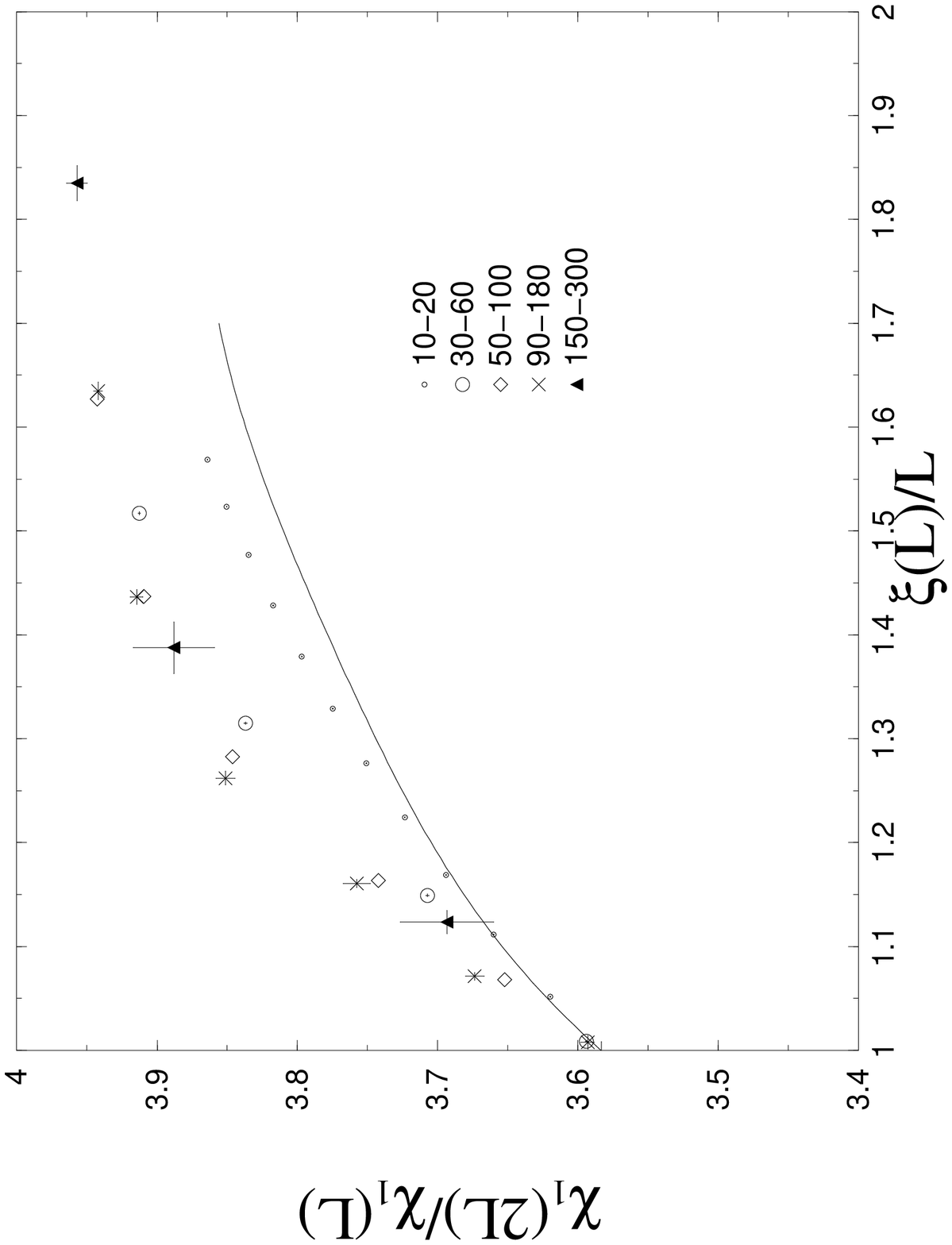,angle=-90,width=1.\linewidth}%
{FSS function for the spin-1 susceptibility $\chi_1$.\label{chi1}}%
{FSS function for the spin-1 susceptibility $\chi_1$. Here, we restrict the 
horizontal range to $1\le x \le 2$.\label{chi1bis}}

In Figs.~\ref{xi}, \ref{xibis}
we report the numerical results for the correlation length. 
The continuous line is a fit to the $O(3)$ data, while the points 
refer to the model with $h=0.1$. 
As observed in previous work, there
is indeed very good agreement between the numerical results 
for the two models, but such an agreement disappears for 
$\xi(L)/L \gtapprox 1$, where small discrepancies are observed. 
As it can be seen from Fig.~\ref{xibis}, the icosahedral points 
tend to be above the $O(3)$ curve and, more importantly, 
the discrepancy tends to increase with $L$: the points 
with $L=10$--20 are systematically below the points 
with $L=50$--100. 

In Figs.~\ref{xi}, \ref{xibis} the difference in behavior 
between the two models is quite small and not totally convincing.
Better evidence is obtained from the results for the susceptibilities,
since in this case the statistical errors are smaller. 
In Figs.~\ref{chi1} and \ref{chi1bis} we report the 
spin-1 susceptibility and in Figs.~\ref{chi3} and \ref{chi4} 
the spin-3 and spin-4 analogues. Again, the numerical results 
for the icosahedral and the $O(3)$ model agree very nicely up to 
$\xi(L)/L \sim 0.8$ -- 1, but then they indicate that the icosahedral 
FSS curve is steeper than the $O(3)$ one. Again, notice that 
the discrepancy between the two models increases with $L$, 
indicating that the observed effect is not due to corrections to scaling, 
i.e. it is not a lattice artifact disappearing in the continuum limit. 
 
\DOUBLEFIGURE[tpb]{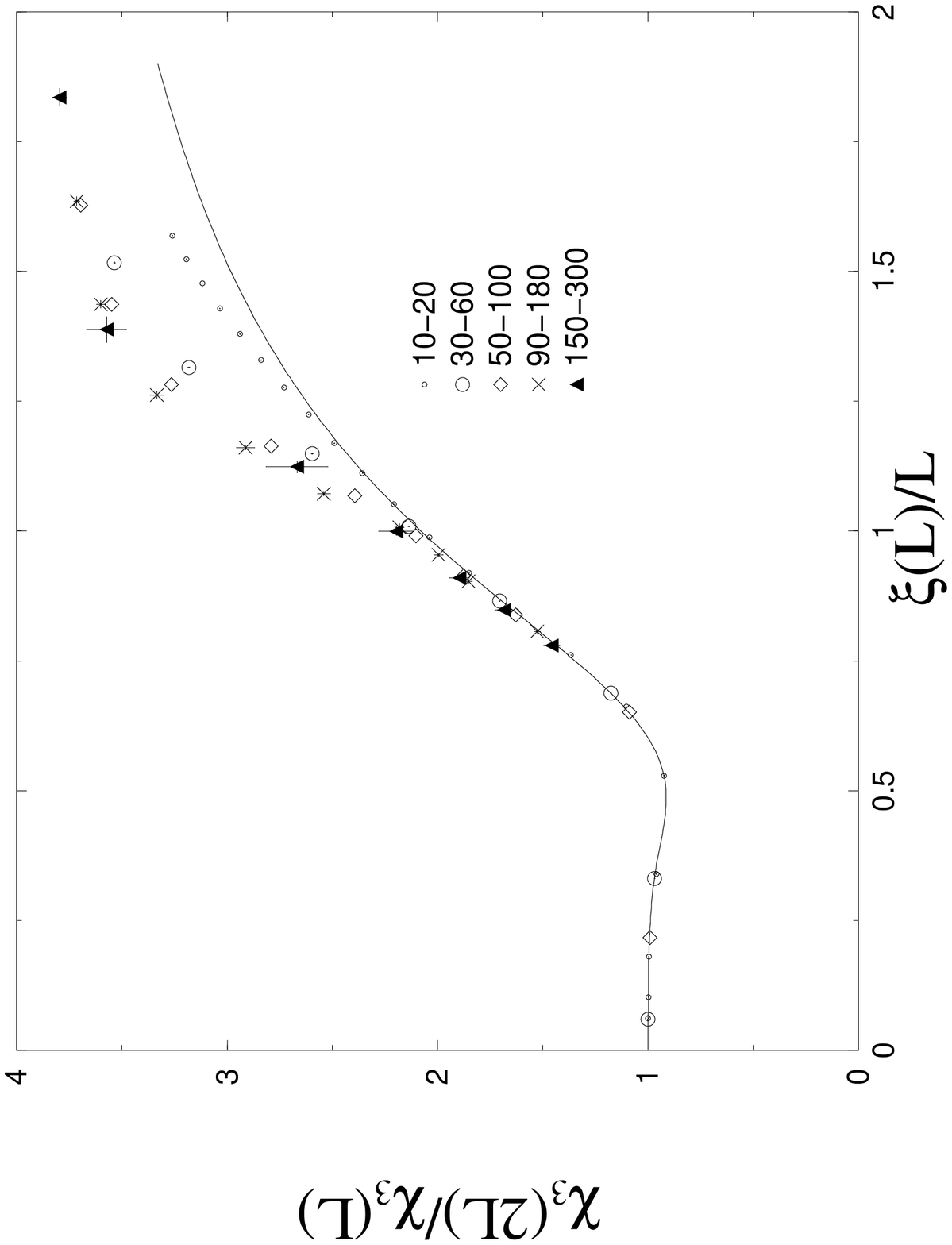,angle=-90,width=1.0\linewidth}%
{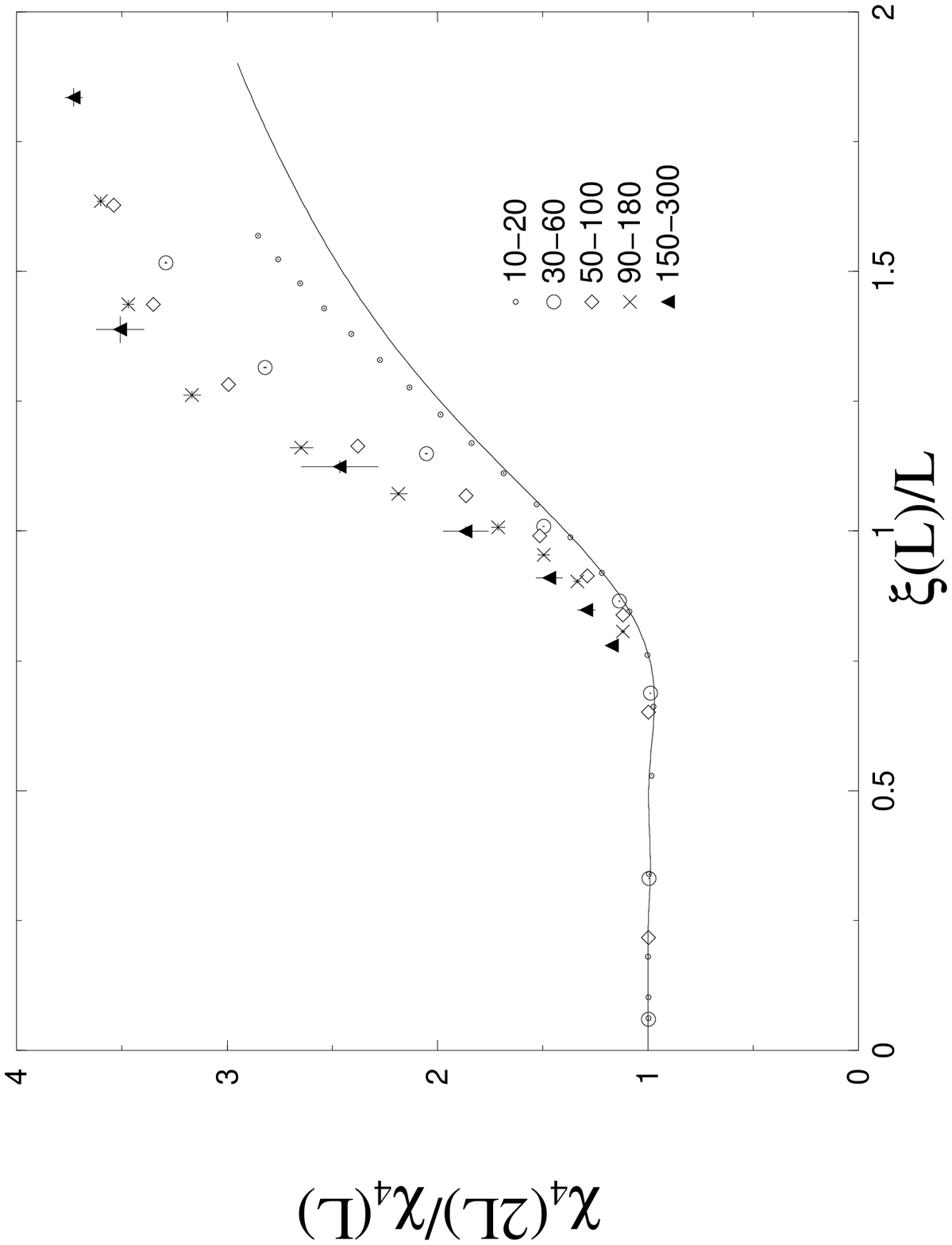,angle=-90,width=1.\linewidth}%
{FSS function for the spin-3 susceptibility $\chi_3$.\label{chi3}}%
{FSS function for the spin-4 susceptibility $\chi_4$.\label{chi4}}

In conclusion, the numerical results show that the icosahedral and the $O(3)$
model belong to different universality classes. Note however that 
discrepancies are observed only for $\xi(L)/L \sim 1$, which corresponds 
to very large values of the infinite-volume correlation length 
(see, e.g. Ref. \cite{CEPS-95}).

We thank Peter Hasenfratz and Ferenc Niedermayer for fruitful 
discussions.

\end{document}